\documentclass{aa}

\usepackage{graphicx}
\usepackage{txfonts}
 \newcommand{\funits}{erg s$^{-1}$ cm$^{-2}$ arcsec$^{-2}$}

\begin{document}
%
   \title{PPAK Integral Field Spectroscopy survey of the Orion Nebula}
   \subtitle{Data release}
   \author{S.F.S\'anchez\thanks{Based on observations collected at the
   Centro Astron\'mico Hispano Alem\'an (CAHA) at Calar Alto, operated jointly
   by the Max-Planck Institut f\"ur Astronomie and the Instituto de Astrof\'\i
   sica de Andaluc\'\i a (CSIC).}
          \inst{1}
          \and
          N.Cardiel\inst{1,2}
          \and
          M.A.W.Verheijen\inst{3}
          \and
          D.Mart\'\i n-Gord\'on\inst{4}
          \and
          J.M.Vilchez\inst{4}
          \and
          J.Alves\inst{1}
          }

 \offprints{S.F.S\'anchez, sanchez@caha.es}
\institute{Centro Astron\'omico Hispano Alem\'an, Calar Alto, CSIC-MPG,
  C/Jes\'us Durb\'an Rem\'on 2-2, E-04004 Almeria, Spain.
  \and
Departamento de Astrof\'\i sica, Facultad de F\'\i sicas, Universidad Complutense de Madrid, 28040 Madrid, Spain. 
  \and
  Kapteyn Astronomical Institute, University of Groningen, PO Box 800, 9700 AV Groningen, The Netherlands.
  \and
  Instituto de Astrof\'\i sica de Andalucia, Camino Bajo de Huetor, S/N, Spain.
}
\date{;}

\abstract{}{We present a low-resolution spectroscopic survey of the Orion
  nebula. The data are released for public use. We show the possible
  applications of this dataset analyzing some of the main properties of the
  nebula.}{We perform an integral field spectroscopy mosaic of an area of
  $\sim$5$\arcmin$$\times$6$\arcmin$ centered on the Trapezium region of the
  nebula, including the ionization front to the south-east. Analysis of the
  line fluxes and line ratios of both the individual and integrated spectra
  allowed us to determine the main characteristics of the ionization
  throughtout the nebula.}{ The final dataset comprises 8182 individual
  spectra, sampled in a circular area of $\sim$2.7$\arcsec$ diameter. The data
  can be downloaded as a single row-stacked spectra fitsfile plus a position
  table or as an interpolated datacube with a final sampling of
  1.5$\arcsec$/pixel.  The integrated spectrum across the field-of-view was
  used to obtain the main integrated properties of the nebula, including the
  electron density and temperature, the dust extinction, the H$\alpha$
  integrated flux (after correcting for dust reddening), and the main
  diagnostic line ratios.  The individual spectra were used to obtain line
  intensity maps of the different detected lines. These maps were used to
  study the distribution of the ionized hydrogen, the dust extinction, the
  electron density and temperature, and the helium and oxygen abundance. All
  of them show a considerable degree of structure as already shown in previous
  studies. In particular, there is a tight relation between the helium and
  oxygen abundances and the ionization structure that cannot be explained by
  case B recombination theory.  Simple arguments like partial ionization and
  dust mixed with the emitting gas may explain these relations. However a more
  detailed modeling is required, for which we provide the dataset.}{}


 \maketitle

\section{Introduction}

The Orion nebula is the brightest and best studied HII region in the sky. It
has been used for decades as a fundamental laboratory in the study of the star
formation regions, ionization processes and helium and heavy element
enrichment. However, despite the large number of studies on this target much
remains unknown about it, even from optical studies. The more we know, the
more complex it seems to be.

Several spectroscopic surveys have tried to characterize its spectroscopic
properties, taking spectra at different ``representative'' locations (e.g.,
\cite{kale76}; \cite{bald91}; \cite{este98}; \cite{este04}). In particular,
Osterbrock et al.  (1990) and Osterbrock et al. (1992) presented a compilation
of high and low dispersion deep spectra of the central bright region in the
optical-NIR region (3000-11000\AA). They measured 225 emission lines, 88 at
the wavelength range of our IFS data. Using the relative intensities of these
lines they derived the relative abundances of several elements, the electron
temperature (T=9000 K) and density ($N_{\rm e}$=4$\times$10$^3$ cm$^{-3}$),
and the extinction ($A_V\sim$1.08 mag, derived from the H$\alpha$/H$\beta$
ratio).  Such properties are frequently compared with those of distant H II
regions in our Galaxy and extragalactic ones (e.g., \cite{page92};
\cite{ferl01}).  Due to their distance most of these regions are poorly
resolved. However their integrated properties are compared with those of
particular resolved areas in the Orion nebula.

Very little effort has been made to study the distribution of physical
properties across the nebula, except for some particular areas
(e.g.\cite{bald91}), or their integrated values in large aperture areas. Pogge
et al. (1992) performed Fabry-Perot (FP) imaging spectrosphotometry on an area
of $\sim$6$\arcmin \times$6$\arcmin$ centered in the Trapezium. Their FP data
covered the brightest emission lines in the nebula: H$\beta$,
[OIII]$\lambda$5007, H$\alpha$, [NII] $\lambda$6548,6583,
[SII]$\lambda$6716,6731 and HeI$\lambda$6678. They derived some of the average
spectroscopic parameters of the core of the nebula, including the integrated
H$\alpha$ flux, the average extinction, the average ionization line ratios
([OIII]/H$\beta$, [NII]/H$\alpha$ and [SII]/H$\alpha$), the He$^+$/H$^+$ ionic
abundance ratio, and the average electron density.  They derived the first
realiable maps of the distribution of these properties across the nebula,
which show a considerable degree of structure.  They found significant
differences with some of the previously reported values that correspond to
particular regions in the nebula (e.g., Osterbrock et al.  1992).  However,
due to the reduced spectroscopic coverage of their dataset it was not possible
to fully characterize the ionization conditions in the nebula.

 \begin{figure*}
   \centering \centering \resizebox{\hsize}{!}
   {\includegraphics[width=\hsize,angle=-90]{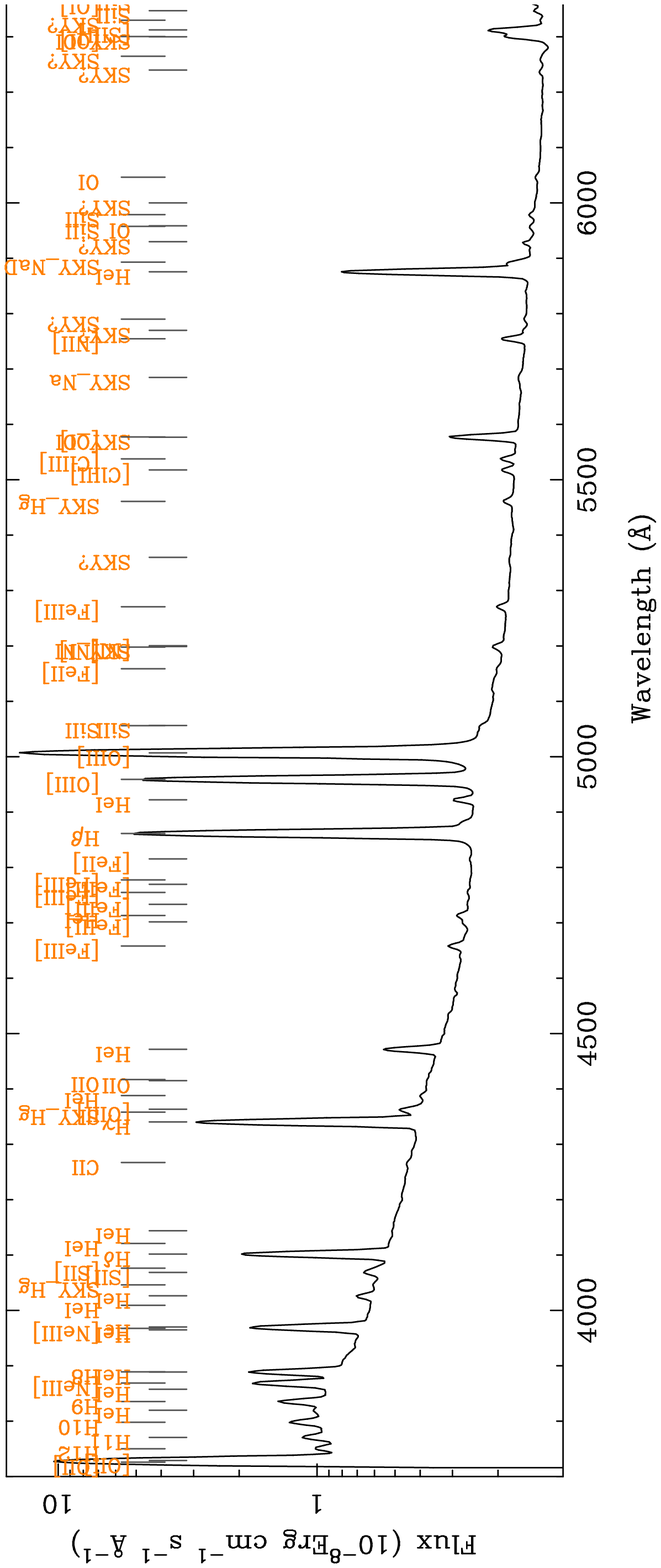}}
 \caption{Integrated spectrum of the Orion Nebula obtained coadding the 8281 individual spectra of
   our IFS dataset over the full field-of-view, plotted on a logarithmic scale.
   The identifications of the detected emission lines, as listed in Table
   \ref{table:1}, are indicated at their corresponding wavelength.}
 \label{fig:id}       
 \end{figure*}

 We present here a low-resolution Integral Field Spectroscopy (IFS) survey of
 a similar field-of-view of the FP observations presented by Pogge et al.
 (1992), but covering the entire optical wavelength range ($\sim$3700-7100
 \AA). The data are obtained, reduced and released for the community. In this
 article we describe the observations and data reduction, indicating the main
 characteristics of the dataset, and we show a few examples of the
 applications of this dataset. To our knowledge this is the largest area ever
 sampled with an IFS unit, showing the capabilities of new tecniques like 3D
 mosaicing to overcome the inherent reduced field-of-view of these
 instruments.
 
 The layout of this article is as follow: In Section 2 we describe the
 observations and data reduction.  Section 3 summarizes the basic properties
 of the dataset. In Section 4 we list examples of the applications of this
 dataset, including the integrated properties of the nebula, the description
 of the line intensity maps of some of the brightest emission lines, a
 distribution of the electron density and temperature, a study of the dust
 extinction, a description of the ionization structure, and a study of the
 helium and heavy element abundances. Section 5 summarizes the conclusions of
 the article.

\section{Observations and data reduction}

Observations were carried out on November 21st, 2004 at the 3.5m telescope of
the Calar Alto observatory with the Potsdam MultiAperture Spectrograph, PMAS,
(\cite{roth05}) in the PPAK mode (\cite{marc04}; \cite{kelz06}). The V300
grating was used, covering a wavelength range between 3700-7100 \AA\ with a
resolution of $\sim$10 \AA\ FWHM, corresponding to $\sim$600 km s$^{-1}$, much
larger than the internal velocities within the nebula. The PPAK fiber bundle
consists of 382 fibers of 2.7'' diameter each (see Fig.5 in Kelz et al.
2006). Of them, 331 fibers (the science fibers) are concentrated in a single
hexagonal bundle covering a field-of-view of 72''$\times$64'', with a filling
factor of $\sim$65\%. The sky is sampled by 36 additional fibers, distributed
in 6 bundles of 6 fibers each, located following a circular distribution
at $\sim$90'' of the center and at the edges of the central hexagon. The
sky-fibers are distributed among the science ones in the pseudo-slit, in order
to have a good characterization of the sky. The remaining 15 fibers are used
for calibration purposes, as described below.

The data were taken as a backup program of the ``Disk Mass Project''
(\cite{marc04}) under bad transparency and non photometric conditions.  A
mosaic of 27 single exposures of 2s each was taken. The exposure time was
selected to avoid saturation of the brightest emission lines, like
H$\alpha$ and [OIII].  The initial exposure was centered in the Trapezium area
of the Orion Nebula.  Consecutive pointings followed a hexagonal pattern,
adjusting the mosaic pointings to the shape of the PPAK science bundle. Each
pointing center lies at 60'' from the previous one. Due to the shape of the
PPAK bundle and by construction of the mosaic, 11 spectra of each pointing,
corresponding to one edge of the hexagon, overlap with the same number of
spectra from the previous pointing. This pattern was selected to maximize the
covered area, allowing enough overlapping to match the different exposures
taken under variable atmospheric conditions.  The offset procedure was carried
out until closing due to risk of rain.  The complete mosaic was completed in
1.5 hours in total, mostly due to the readout time of the PMAS CCD camera
($\sim$1.5 min in the setup used).

\begin{table*}
\caption{List of detected emission lines, uncorrected for extinction.}             
\label{table:1}      
\centering                          
\begin{tabular}{llrc|llrc}        
\hline\hline                 
Wavelength & Wavelength & Flux$^1$& Line Id. & Wavelength & Wavelength & Flux& Line Id. \\
Lab. (\AA) & Obs. (\AA) & 
&&
Lab. (\AA) & Obs. (\AA) & 
\\
\hline   
3726.03$^{\bf *}$& 3724.94 & 35219.70& [OII] & 5006.84$^{\bf *}$& 5006.99 & 143002.31& [OIII] \\
3728.82$^{\bf *}$& 3730.08 & 40670.09& [OII] & 5056.02& 5055.80 &   35.09& SiII \\
3750.15& 3751.09 &  676.63& H12              & 5056.35& 5056.35 &   39.79& SiII \\
3770.63& 3771.25 & 1574.32& H11              & 5158.81& 5157.92 &   52.29& [FeII] \\
3797.90& 3798.18 & 2550.31& H10              & 5197.90& 5195.79 &   25.86& [NI] \\
3819.64& 3819.54 &  876.94& HeI3819          & 5199.00& 5199.49 &   76.77&  SKY\_NI \\
3835.39& 3835.15 & 3391.71& H9               & 5200.26& 5201.02 &  142.51& [NI] \\
3857.53& 3853.55 &  561.27& HeI              & 5270.40& 5270.64 &  171.06& [FeIII] \\
3868.75& 3868.59 & 8266.59& [NeIII]          & 5360.00& 5359.94 &    6.06& SKY? \\
3888.65& 3887.97 & 4416.27& HeI              & 5461.00& 5461.07 &  215.29& SKY\_HgI \\
3889.05& 3890.22 & 3957.53& H8               & 5517.71& 5517.71 &  284.96& [ClIII] \\
3964.73& 3967.69 & 3634.16& HeI              & 5537.88& 5537.89 &  281.24& [ClIII] \\
3967.46& 3968.09 &  642.14& [NeIII]          & 5577.00& 5576.40 &  789.97& SKY\_OI \\
3970.07& 3969.20 & 5478.41& H$\epsilon$      & 5577.31& 5578.10 &  791.73& [OI] \\
4009.27& 4007.41 &  318.24& HeI              & 5685.00& 5684.97 &  118.17& SKY\_NaI \\
4026.21& 4026.00 &  722.02& HeI              & 5754.64$^{\bf *}$& 5754.66 & 400.01& [NII] \\
4046.00& 4046.02 &   32.09& SKY\_HgI         & 5770.00& 5769.91 &   15.66& SKY? \\
4068.60& 4068.39 &  458.02& [SII]            & 5790.00& 5790.04 &   1.16& SKY? \\
4076.35& 4074.88 &  159.72& [SII]            & 5875.62& 5875.56 & 7538.90& HeI \\
4101.74& 4101.56 & 12021.26& H$\delta$       & 5893.00& 5892.61 &  348.91& SKY\_NaI D \\
4120.86& 4121.23 &   53.23& HeI              & 5930.00& 5927.95 &  123.02& SKY? \\
4143.76& 4148.23 &   23.02& HeI              & 5957.61& 5957.11 &   41.97& SiII \\
4267.15& 4265.12 &   85.48& CII              & 5958.58& 5957.49 &   12.78& OI \\
4340.47& 4340.32 & 23860.65& H$\gamma$       & 5978.97& 5978.80 &   81.35& SiII \\
4358.00& 4364.15 &  113.00& SKY\_HgI         & 6000.00& 6000.26 &   12.91& SKY? \\
4363.21$^{\bf *}$& 4366.76 & 831.27& [OIII]  & 6046.40& 6046.57 &   41.92& OI \\
4387.93& 4379.43 &  125.13& HeI              & 6240.00& 6239.95 &   60.97& SKY? \\
4414.91& 4471.37 & 1937.47& OII              & 6265.00& 6264.91 &   30.67& SKY? \\
4416.98& 4418.47 &   62.93& OII              & 6300.00& 6299.33 &  297.01& SKY\_OI \\
4471.50& 4412.41 &   97.99& HeI              & 6300.30& 6301.39 &  362.98& [OI] \\
4658.10& 4658.22 &  379.59& [FeIII]          & 6312.10& 6312.27 &  950.13& [SIII] \\
4701.62& 4701.26 &  123.43& [FeIII]          & 6330.00& 6329.99 &   46.28& SKY? \\
4713.20& 4713.36 &  279.38& HeI              & 6347.09& 6347.30 &  141.11& SiII \\
4733.93& 4735.80 &   40.15& [FeIII]          & 6363.78& 6363.72 &  208.24& [OI] \\
4754.83& 4755.52 &   62.28& [FeIII]          & 6370.36& 6370.53 &  107.90& SiII \\
4769.60& 4775.10 &   62.48& [FeIII]          & 6548.03$^{\bf *}$& 6547.63 & 10743.11& [NII] \\
4777.88& 4783.04 &   29.99& [FeIII]          & 6562.82$^{\bf *}$& 6562.87 & 171573.70& H$\alpha$ \\
4815.55& 4815.82 &   82.38& [FeII]           & 6583.41$^{\bf *}$& 6583.52 & 32494.66& [NII] \\
4861.33$^{\bf *}$& 4861.41 & 50768.69& H$\beta$ & 6678.15$^{\bf *}$& 6678.37 & 1907.79& HeI6678 \\
4921.93& 4922.20 & 201.94& HeI               & 6716.39& 6716.47$^{\bf *}$& 2354.08& [SII] \\
4958.91$^{\bf *}$& 4959.02 & 47337.07& [OIII]& 6730.85$^{\bf *}$& 6731.27 & 3447.13& [SII] \\
\hline                                   
\end{tabular}

$^1$ 10$^{-11}$ Erg cm$^{-2}$ s$^{-1}$

$^{\bf *}$ Emission line used in our analysis of the data.

\end{table*}

Data reduction was performed using R3D (\cite{sanc06}), in combination with
IRAF packages (\cite{iraf})\footnote{IRAF is distributed by the National
  Optical Astronomy Observatories, which are operated by the Association of
  Universities for Research in Astronomy, Inc., under cooperative agreement
  with the National Science Foundation.} and E3D (\cite{sanc04}). The
reduction consists of the standard steps for fiber-based integral-field
spectroscopy. A master bias frame was created by averaging all the bias frames
observed during the night and subtracted from the science frames. The location
of the spectra in the CCD was determined using a continuum illuminated
exposure taken before the science exposures. Each spectrum was extracted from
the science frames by coadding the flux within an aperture of 5 pixels around
this location along the cross-dispersion axis for each pixel in the dispersion
axis and stored in a row-staked-spectrum file RSS (\cite{sanc04}). Wavelength
calibration was performed using a He lamp exposure obtained from archive data,
and corrected for distortions using ThAr exposures obtained simultaneously to
the science exposures through the calibration fibers (indicated above).
Differences in the fiber-to-fiber transmission throughput were corrected by
comparing the wavelength-calibrated RSS science frames with the corresponding
continuum illuminated ones. The nature of the dataset prevents us from
performing any sky subtraction. However, due to the strength of the detected
emission lines and the short exposure time, sky contamination is not an
important issue for most of the foreseeable science cases.

Since the data were taken under bad weather conditions, no attempt was made to
perform a flux calibration on the basis of the standard stars observed during
the night. On the contrary, flux calibration was performed following a
two-step procedure. First, we carried out a flux calibration using a standard
star observed with a similar instrumental setup on a previous night. This
procedure corrects, to a first order, for the instrumental transmission. Then,
we used multi-band photometric data available in the literature for the
brightest star in the Trapezium, $\Theta^1$ Ori C (\cite{lee68};
\cite{vitr00}) to perform an absolute flux calibration by comparing with the
un-calibrated spectrum of this object extracted from the central pointing of
the mosaic. An additional correction was performed to take into account flux
loses in the un-calibrated spectrum of the object due to differential
atmospheric refraction. This correction was determined by comparing the
emission line intensities with those derived by Baldwin et al.  (1991) in the
spatial region (O'Dell, private communications). This procedure ensures a flux
calibration and sky extinction correction of this central pointing.

After reducing each individual pointing we built a single RSS file for the
mosaic following an iterative procedure. The spectra of each pointing was
scaled to those of the previous one by the average ratio of the emission of
H$\beta$ in the overlapping spectra. To address the possible variable
extinction during the observations, the average ratio between these scaled
spectra and the previous ones was determined. We fitted this
wavelength-dependent ratio to a low order polynomial function and divided all
the spectra in the new pointing by the resulting wavelength dependent scale.
In most of the cases this second correction has little effect on the data,
indicating that the sky extinction did not change too strongly during the
observations, prehaps because they comprise only a limited amount of time
(1.5h).  The overlapping spectra were then replaced by the average between the
previous pointing and the new rescaled spectra. The resulting spectra were
included in the final RSS file, updating the corresponding position table.
The process starts in the central pointing, that is flux calibrated and sky
extinction corrected as explained before, ensuring an homogeneous flux
calibration and sky extinction correction for all the dataset. The procedure
was repeated until the last pointing was included, ending with a final set of
8182 individual spectra and its corresponding position table. In addition , we
created a datacube with 1.5''/pixel sampling by interpolating this RSS frame
using E3D (\cite{sanc04}).

\section{Results}

The final data set consists of 8182 flux calibrated spectra sampling a
circular area of $\sim$2.7$\arcsec$ diameter each and covering a contiguous
field-of-view of $\sim$5$\arcmin$$\times$6$\arcmin$ centered approximately on
the Trapezium area of the Orion nebula. Each spectrum covers the wavelength
range between $\sim$3700-7100 \AA, which includes the most prominent
recombination and collisionally excited emission lines, from
[OII]$\lambda$3727,3729 to [SII]$\lambda$6716,6731 (see Table 1). Near the
lower wavelength range, the [OII] flux was measured without problem in all
fibers. The faintest detected emission lines, up to 5$\sigma$, in the
individual spectra have surface brightness of the order of 10$^{-14}$ \funits.
It is out of the scope of this article to extract all the information
potentially available in this huge dataset. We consider here a few
relevant properties of the Orion nebula using these data. Howeverm we are
releasing the full dataset in order to make if freely available to the
community on the project webpage \footnote{\tt
  http://www.caha.es/sanchez/orion/}.

\section{Applications}

\subsection{Integrated properties of the Nebula}
\label{sec:inte}

In this section we describe some of the more interesting spectroscopic
properties of the nebula derived by analyzing the integrated spectrum over the
full field-of-view of our IFS dataset. Contrary to previous studies that
attempt to describe the average ionization conditions in the nebula by
analyzing individual spectra taken at different locations, we could, for the
first time, describe the real average spectroscopic parameters.

\subsubsection{Integrated spectrum of the Nebula}
\label{sec:inte_spec}

Figure \ref{fig:id} shows the integrated spectrum of the Orion Nebula obtained
by coadding the 8281 individual spectra of our IFS dataset, plotted on a
logarithmic scale. 82 emission lines are detected in the integrated spectrum.
They were identified using the list of 88 emission lines detected by
Osterbrock et al. (1992) using deep spectroscopy in the same wavelength range,
in combination with the list of known sky emission lines at Calar Alto
(S\'anchez, in prep.).  The observed wavelength and integrated flux of each
emission line were measured using FIT3D (\cite{sanc06b}), removing the
background continuum and deblending different emission lines when needed (and
possible). Table \ref{table:1} lists the detected emission lines, including,
for each line, the laboratory wavelength (\cite{oster92}), the observed
wavelength, the integrated measured flux and the line identification. The
final effective resolution of our integrated spectrum was FWHM$\sim$10\AA,
measured in the brightest detected lines. Therefore, any emission line at
distances less than $\sim$ 3\AA\ from its adjacent neighbors is blended, and
the listed integrated flux must be taken with care.

There is a considerable amount of blue continuum emission in the integrated
spectrum shown in Fig. \ref{fig:id}, whose intensity level ranges from
$\sim$10$^{-8}$ erg com$^{-2}$ s$^{-1}$ \AA$^{-1}$ in the ultraviolet to a few
10$^{-9}$ erg com$^{-2}$ s$^{-1}$ \AA$^{-1}$ in the reddest wavelength ranges.
This continuum is partially due to the stars sampled by our IFS data, in
particular the bright members of the Trapezium. However, it is well known
(e.g., Pogge et al. 1992) that the Orion nebula presents a diffuse continuum
emission, whose origin is associated with the light from the stars within the
ionized gas that is scattered by dust. Unfortunally, the measured continuum is
strongly affected by the fact that we could not subtract the sky emission.
Therefore this issue cannot be addressed propertly with the current IFS data.


\subsubsection{Average dust extinction}
\label{sec:inte_AV}

The H I Balmer series are normally used to determine the extinction. One must
compare the observed line ratios with the expected theoretical values. These
lines are known to be emitted as a consequence of recombination, and the
theory for them has been worked out in considerable detail (\cite{humm87};
\cite{oster89}). Case B (optically thick in all the Lyman lines) is the best
simple approximation to describe the real physical conditions in the
ionization of the gas in the Orion nebula (\cite{oster88}; \cite{oster92};
\cite{ferl01}). The main deviation of the real nebula from the theory is
probably the absorption of Lyman-line photons by dust mixed with the emitting
gas, whose observable effects are at the level of a few percent in the
H$\alpha$/H$\beta$ ratio (\cite{cota88}; \cite{bald91}). Table \ref{table:2}
lists the observed and expected ratios between the H I Balmer series lines
listed in Table \ref{table:1}. They show a good agreement within the errors,
particularly the ratio for the 5 brightest emission lines (whose errors are
estimated to be lower than $\sim$10\%), indicating the presence of little
dust extinction on average.

 \begin{figure*}
   \centering \centering \resizebox{\hsize}{!}
  {\includegraphics[width=7.8cm,angle=-90]{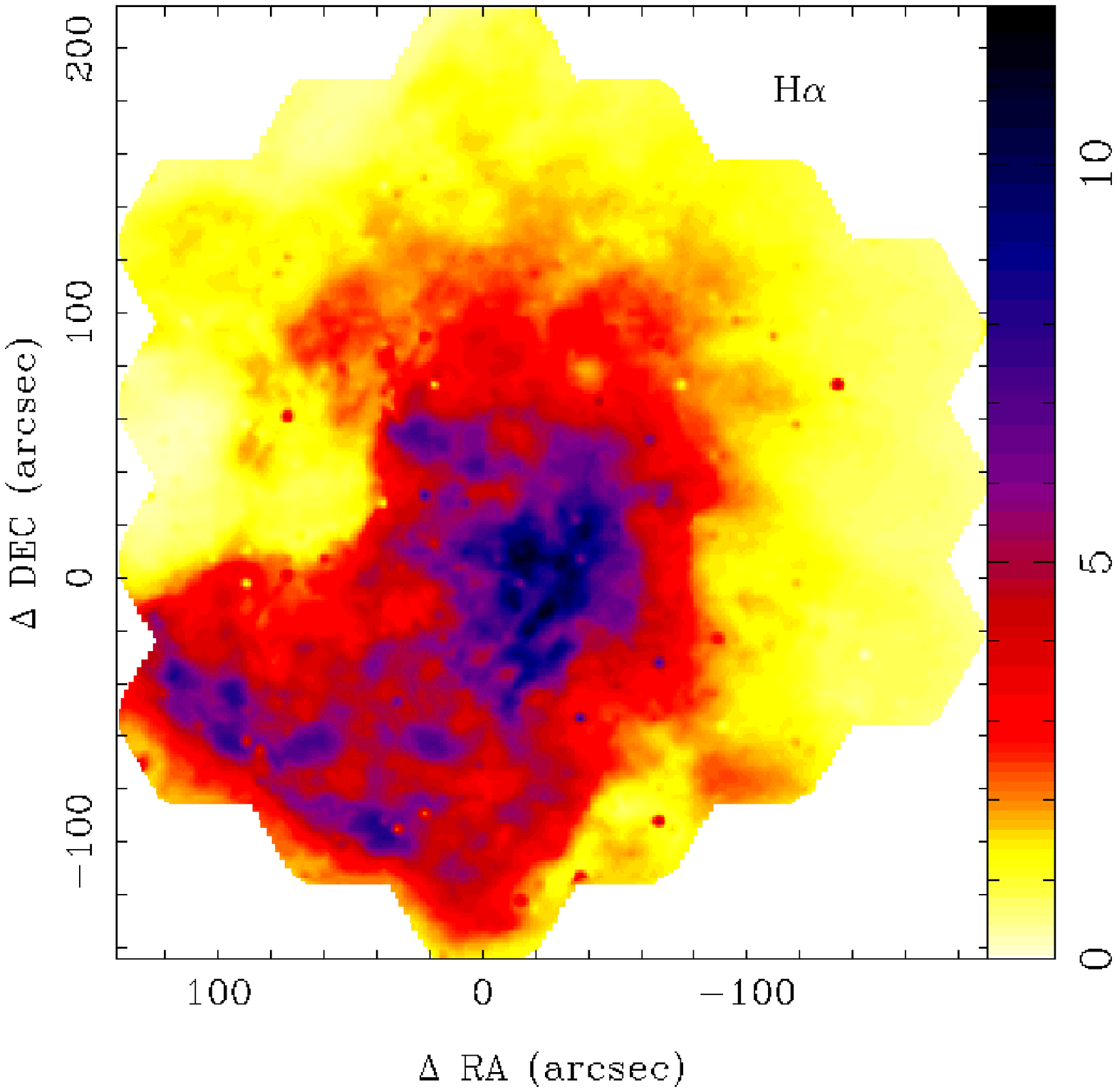}\includegraphics[width=7.8cm,angle=-90]{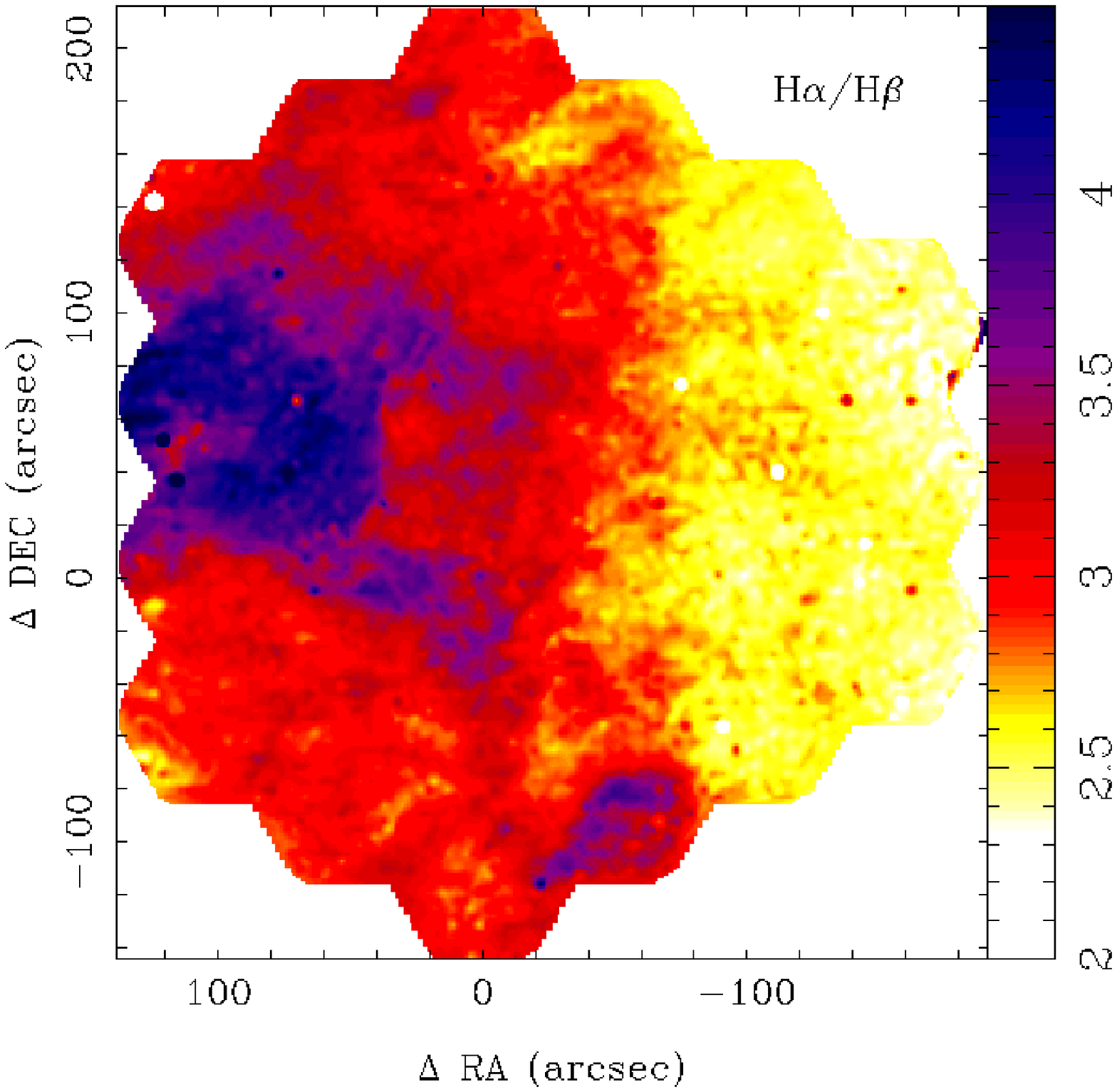}}
 \caption{Left: observed H$\alpha$ line flux map in units of 10$^{-12}$
   erg s$^{-1}$ cm$^{2}$ arcsec$^{-2}$ obtained by fitting a single Gaussian
   function to the H$\alpha$ emission line for each single spectrum in the
   dataset. The intensity map is then reconstructed by interpolating the
   recovered flux at the location of each spectrum, as described in the text.
   Right: observed H$\alpha$/H$\beta$ line ratio map.  No correction for dust
   extinction was applied to any maps. }
 \label{fig:Ha}       
 \end{figure*}

 An average extinction of $A_V$=0.75 mag is found assuming an intrinsic value
 of H$\alpha$/H$\beta$=2.86, corresponding to case B recombination for T=9000
 K (\cite{humm87}; \cite{pogg92}; \cite{oster92}), and using the Orion
 extinction law of Cardelli, Clayton \& Mathis (1989), with a ratio of total
 to selective extinction of $R_V=$5.5 (\cite{math81}). This average extinction
 is smaller than the value reported by Pogge et al. (1992). However, a
 detailed comparison of the field-of-view covered by their FP data and our IFS
 ones shows that their data are approximately centered $\sim$60$\arcsec$ more
 towards the east, covering a larger area of the Orion ``Dark Bay'', an area
 with dust extinctions up to $A_V$$\sim$2 mag.  There are better line ratios
 to estimate the extinction (\cite{oster92}), since, as we noted before, the
 intrinsic H$\alpha$/H$\beta$ ratio in the real nebula is a few percent
 smaller than the case B value (e.g, \cite{bald91}).  We will discuss its
 effect in further sections.

\begin{table}
\caption{Observed Balmer line ratios, uncorrected for extinction, compared
 with the theoretical values obtained from Osterbrock et al. (1992)}             
\label{table:2}      
\centering                          
\begin{tabular}{lrl}        
\hline\hline                 
Line ratio & Obs. Value & Exp. Value \\
\\
\hline   
H$\alpha$/H$\beta$       &3.380         &2.86    \\
H$\gamma$/H$\beta$       &0.470         &0.486   \\                        
H$\delta$/H$\beta$       &0.237         &0.259   \\                      
H$\epsilon$/H$\beta$$^*$ &0.108         &0.159   \\
H8/H$\beta$$^{**}$       &0.078         &0.105   \\
H9/H$\beta$              &0.067         &0.0731  \\
H10/H$\beta$             &0.050         &0.0530  \\
H11/H$\beta$             &0.031         &0.0398  \\
H12/H$\beta$             &0.013         &0.0307  \\
\hline                                   
\end{tabular}

$^*$ Blended with  [NeIII]

$^{**}$ Blended with  HeI

\end{table}

\subsubsection{Average electron density and temperature}
\label{sec:inte_Ne_T}

The electron density, $N_e$, in the S$^+$ zone of the nebula can be estimated
using the observed [SII]$\lambda$6716,6731 doublet ratio (\cite{oster89}). The
estimated [SII]$\lambda$6716/[SII]$\lambda$6731 ratio based on the integrated
spectrum is 0.683, that can be converted to an electron density, $N_e$.
Although mild, this conversion shows a dependency with the electronic
temperature ($\sim N_e/T^{1/2}$, \cite{oster89}).  Many previous
spectroscopic studies sampling the Orion nebula at different locations
reported a range of temperatures between 8000-10000~K (e.g., \cite{oster92}).
Therefore, the measured ratio could correspond to electron densities in a
range between 2000-4000 cm$^{-3}$. Pogge et al. (1992) reported an average
[SII]$\lambda$6716/[SII]$\lambda$6731 ratio of 0.680, similar to our
measurement. However, since their FP data did not cover any of the usual
temperature indicators, they converted it to an electron density of $N_e=$2150
cm$^{-3}$ assuming an average temperature of 9000 K (\cite{bald91}).  Our IFS
spectroscopic data do cover at least two of the classical indicators with
enough signal-to-noise, the [OIII] ratio:

$$\frac{I(\lambda4959)+I(\lambda5007)}{I(\lambda4363)}=228.97$$

and the [NII] ratio:

$$\frac{I(\lambda6548)+I(\lambda6583)}{I(\lambda5755)}=108.09$$

Both ratios are remarkably similar to the values reported by Osterbrock et al.
(1992), for the center of the Nebula, despite the evident variable conditions
of the ionization across the nebula. 

The conversion between both ratios and the temperature, $T$, depends on
$N_e/T^{1/2}$ (\cite{oster89}). Therefore, it is possible to derive the
temperature using both ratios and the ratio between the [SII] doublet. Once
the temperature is obtained, it is possible to derive the electron density
too.  We found an electronic density of $N_e=$2104 cm$^{-3}$ (2019 cm$^{-3}$)
and a temperature of $T$=9682 K ($T$=8922 K) for the [OIII] ([NII]) ratio. To
derive the electron density from the [SII] lines ratio and the temperature
derived from the [OIII] lines ratios (T$_3$), we transformed this temperature
to T$_2$ using the relation between both quantities presented by Pilyugin et
al. (2006).  Since the [OIII]$\lambda$4363 may be affected by an unsubtracted
night sky HgI emission line at $\lambda$=4358\AA\ we consider the estimation
based on the [NII] ratio more reliable. Although the lines were not corrected
by extinction, its effect on the derived temperature is negligible (even for
$A_V>$2 mag). The average oxygen abundance, 12+log(O/H) $\sim$ 8.6, derived
from the fluxes of the integrated spectrum quoted in Table 1 is close to the
solar value.

 \begin{figure*}
   \centering \centering \resizebox{\hsize}{!}
   {\includegraphics[width=7.8cm,angle=-90]{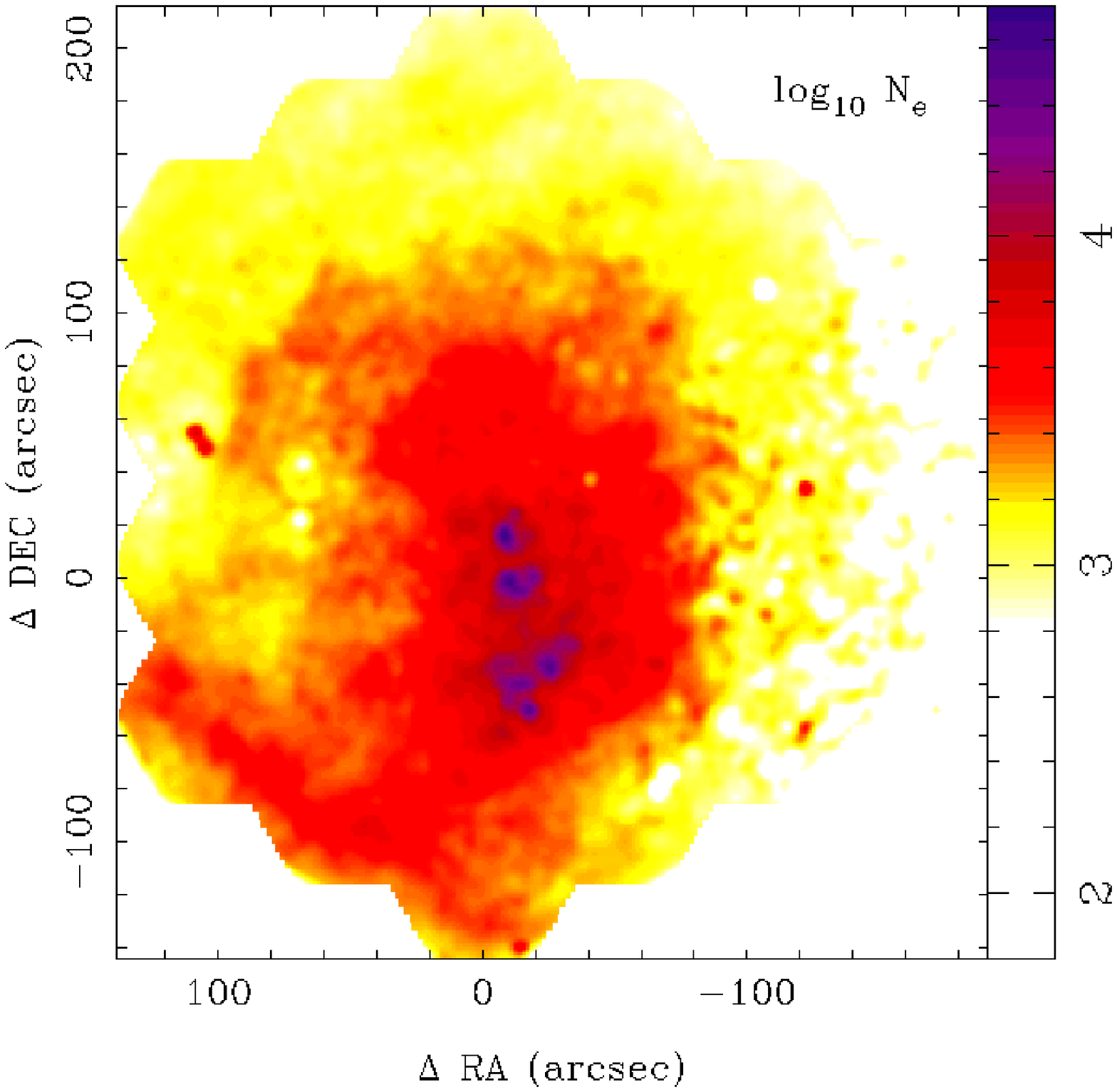}\includegraphics[width=7.8cm,angle=-90]{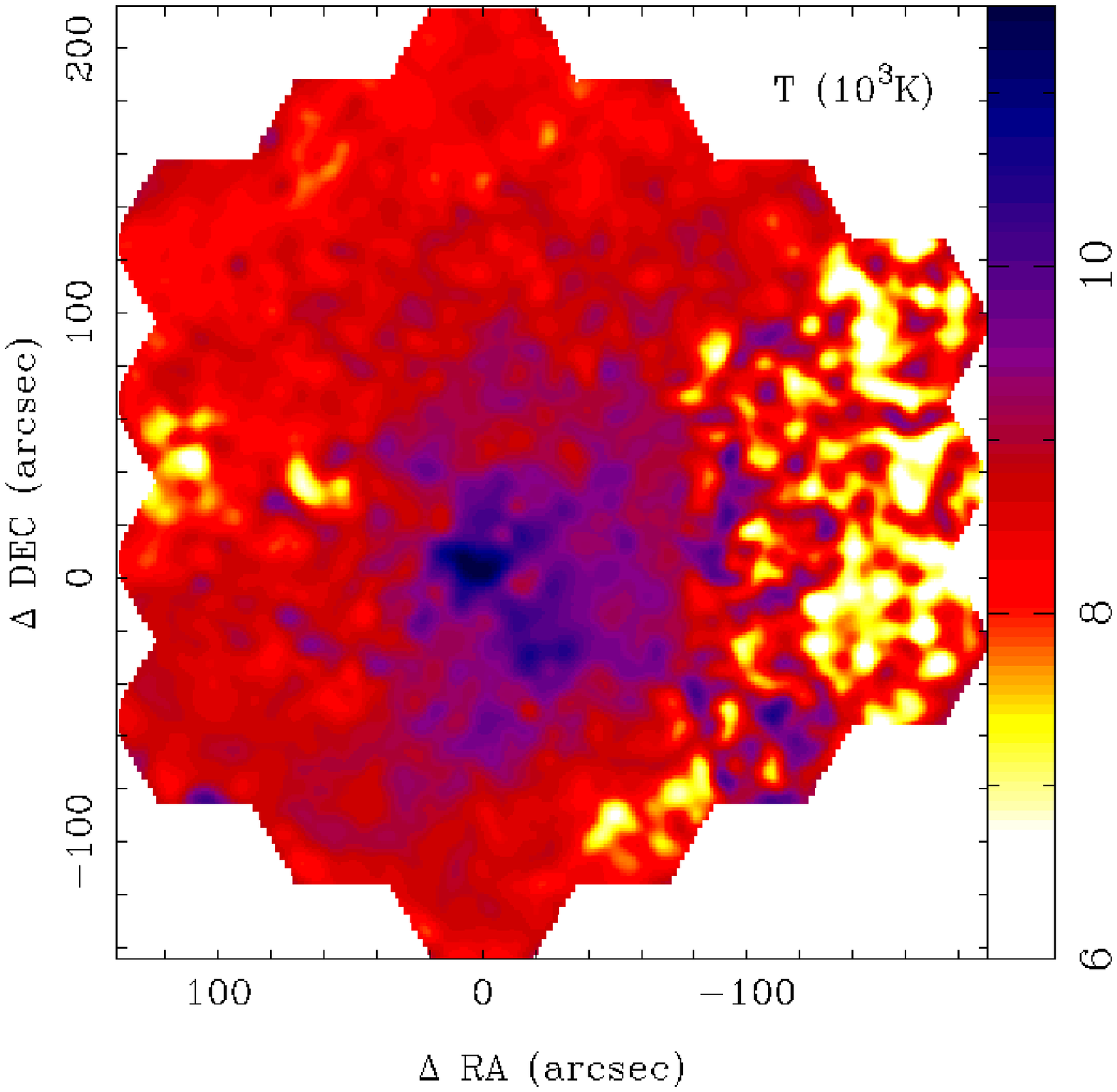}}
 \caption{Left: Electron density map, $N_e$ in $cm^{-3}$, derived from the
   [SII]$\lambda$6716/[SII]$\lambda$6731 line ratio, represented in a
   logarithm scale. Right: Electron temperature map in units of 10$^3$K,
   derived from the ([NII]$\lambda$6548+[NII]$\lambda$6583)/[NII]$\lambda$5755
   line ratio.}
 \label{fig:Ne_T}       
 \end{figure*}

\subsubsection{Average ionization conditions}
\label{sec:inte_OIII}

It is widely accepted that the ionization of the Orion nebula is dominated by
photoionization by the star $\Theta^1$ Ori C. The derived values for the
classical diagnostic line ratios, [OIII]$\lambda$5007/H$\beta$=2.82,
[NII]$\lambda$6583/H$\alpha$=0.19 and [SII]$\lambda$6716+31/H$\alpha$=0.034, are
consistent with this picture, being the typical values for H II regions
ionized by OB stars (\cite{oster89}). As already noticed by Pogge et al.
(1992), the integrated H$\alpha$ luminosity, 7.6$\times$10$^{37}$ erg s$^{-1}$
(extinction corrected), and the electron density indicate too that Orion is a
rather dense, low-luminosity H II region (\cite{kenn89}). Indeed, this
H$\alpha$ luminosity agrees with the nebula ionization dominated by a single
O 5 star (\cite{pogg92}; \cite{humm87}). A summary of the average properties
of the nebula derived from our integrated spectrum is listed in Table
\ref{table:3}.

\begin{table}
\caption{Integrated Spectroscopic properties of the nebula.}             
\label{table:3}      
\centering                          
\begin{tabular}{ll}        
\hline\hline                 
Parameter & Value \\
\\
\hline   
$A_V$   & 0.75 (mag)\\
$N_e$   & 2019 (cm$^{-3}$)\\
T$_2^*$     & 8922 (K)\\
$[$OIII$]$$\lambda$5007/H$\beta$ & 2.820\\
$[$NII$]$$\lambda$6583/H$\alpha$ & 0.190\\
$[$SII$]$$\lambda$6716+31/H$\alpha$ & 0.034\\
\hline                                   
\end{tabular}

$^*$ Electron temperature derived from double ionized ions, like N$^{++}$, using
the formulae described in Osterbrock (1989).
\end{table}

\subsection{Spatial distribution of the properties of the Nebula}
\label{spatial}

While in previous sections we focused on the integrated properties of the
nebula, in this section we describe the spatial distribution of these
properties along the nebula. We use the 8182 individual spectra as independent
probes of these properties at different locations.

\subsubsection{Line intensity maps}
\label{line_map}

The ionized gas in the Orion nebula exhibits a rather complex structure. Due
to that, it has been targeted by narrow-band imaging using the HST and
groundbased telescopes in order to disentangle the structure of its ionized
gas (e.g., Wen and O'Dell 1995; \cite{odel00}; \cite{odel03}; \cite{rubi03}
and references therein).  In many cases these narrow-band images catch more
than one single line (e.g., H$\alpha$, the [NII]$\lambda$6548,6583 doublet or
the [SII]$\lambda$6716,6731 doublet), reducing its use to study the basic
parameters of the ionized gas (as listed in Table \ref{table:3}). As we quoted
before, Pogge et al. (1992) presented FP observations that allowed them to
disentangle some of the main diagnostic emission lines. Following them we have
obtained line emission maps for the same emission lines. We have also derived
line emission maps for [NII]$\lambda$5755, [OIII]$\lambda$4363 and
[OII]$\lambda$3727,29, used as temperature, oxygen abundance and ionization
indicators in combination with the lines quoted before. All the analyzed lines
are listed in Table \ref{table:1}, labeled with an asterisk.

To derive the emission line maps each spectrum in the dataset was fitted with
a single Gaussian function per emission line plus a low order polynomial
function to describe the continuum emission again using FIT3D
(\cite{sanc06b}). Instead of fitting the entire wavelength range in a row, we
extracted shorter wavelength ranges for each spectrum that sampled one or a
few of the analyzed emission lines, in order to characterize the continuum
with the most simple polynomial function, and to simplify the fitting
procedure.  In a few cases we had to include some of these lines: e.g.,
[OIII]$\lambda$4363 is blended with a night sky HgI line that must also be
included in the fit. When more than one emission was fitted simultaneously,
their FWHMs were forced to be equal (since the FWHM is dominated by the
spectral resolution), in order to decrease the number of free parameters and
increase the accuracy of the deblending process (when required).  A final
interpolated map with a 1$\arcsec$/pixel scale was derived for each emission
line by interpolating the discrete sample of intensities obtained for each
spectrum (located at a certain position in the sky).  The interpolation was
performed using the routines included in E3D (\cite{sanc04}). The data at the
location of bright stars in the field were masked prior to any interpolation,
in order to decrease the effects of their contamination.

 \begin{figure*}
   \centering \centering \resizebox{\hsize}{!}
   {\includegraphics[width=7.8cm,angle=-90]{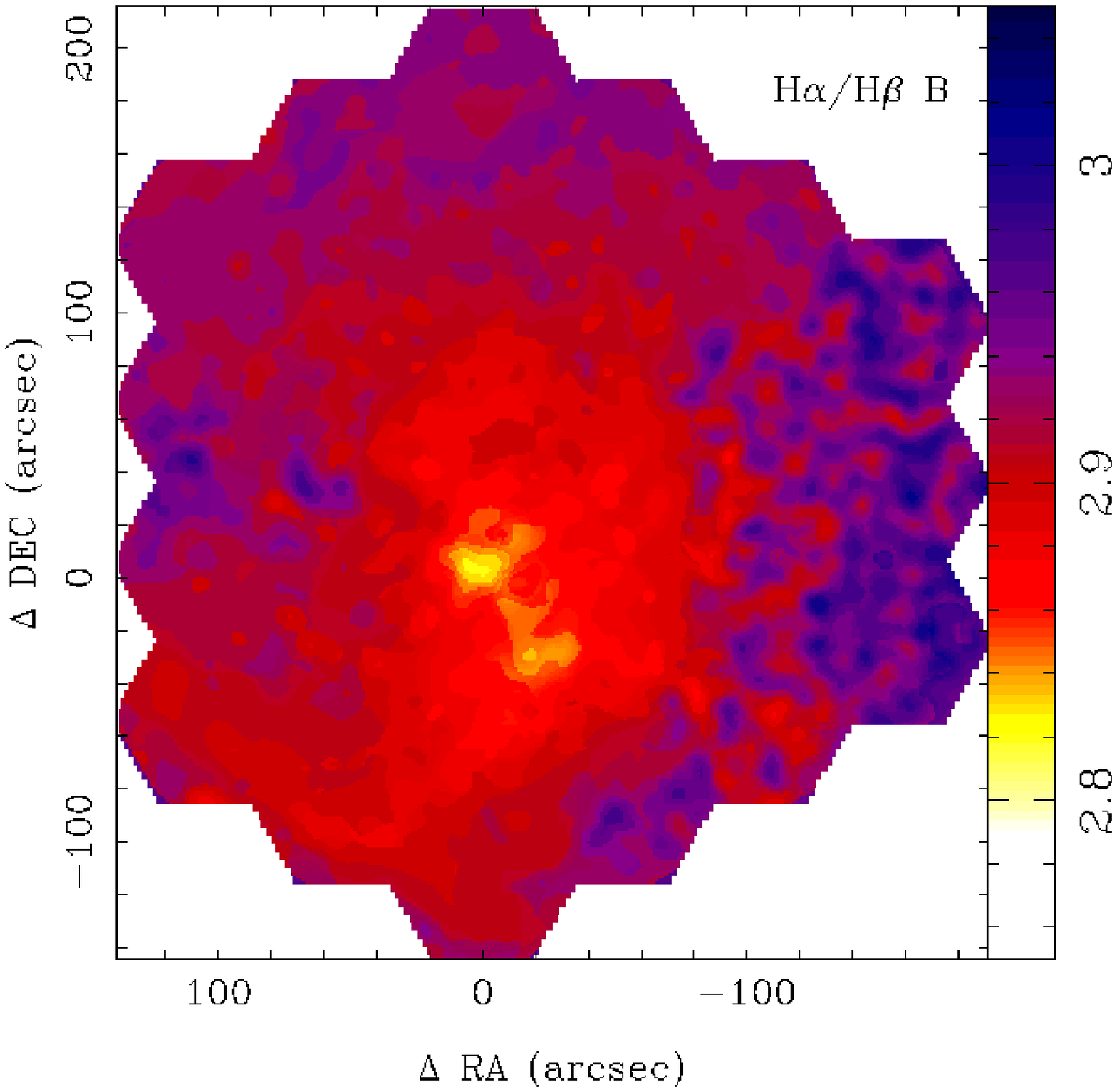}\includegraphics[width=7.8cm,angle=-90]{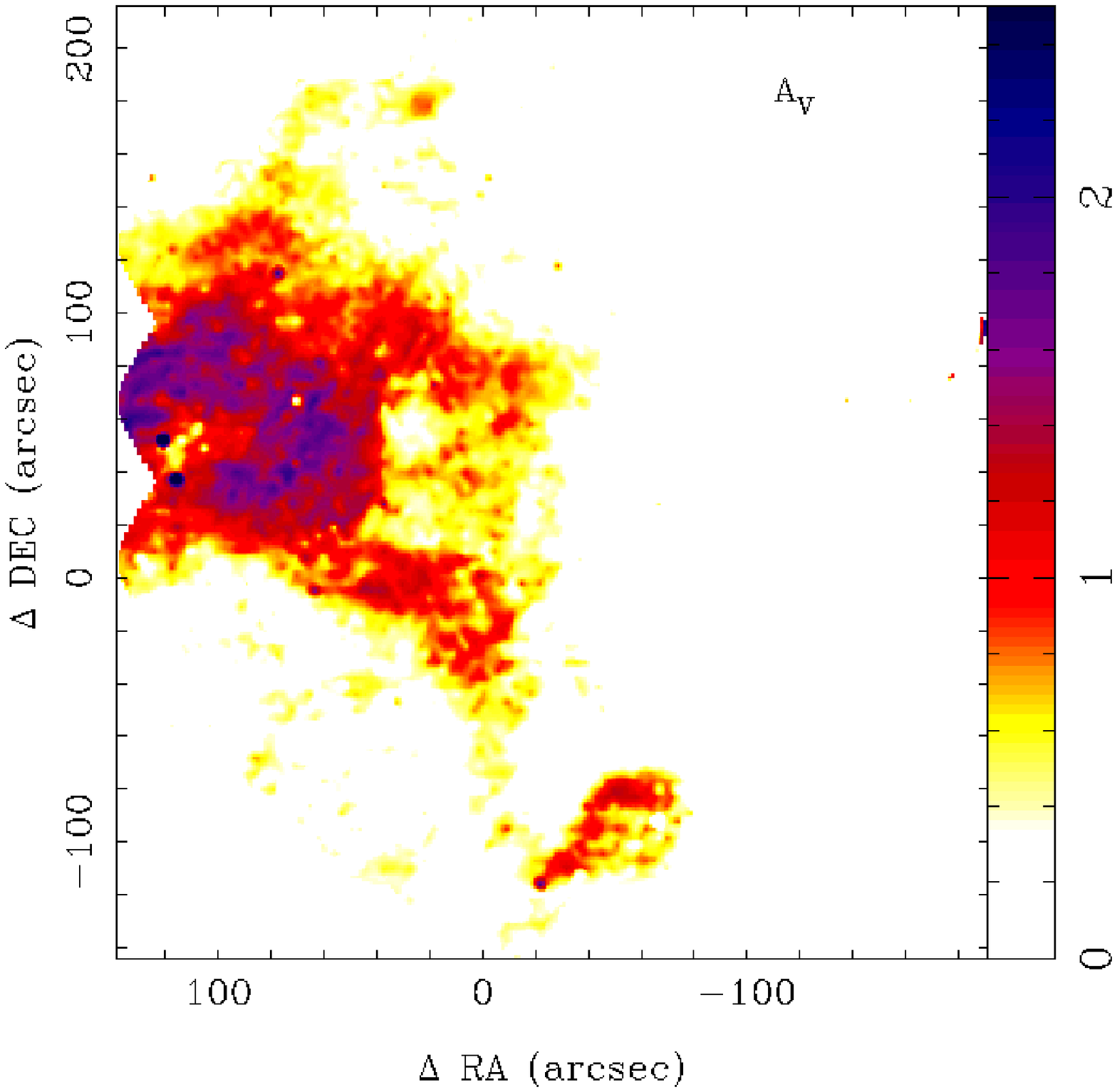}}
 \caption{Left: H$\alpha$/H$\beta$ line ratio map calculated for case B
   recombination using the electron density, $N_e$, and temperature maps
   described above (Fig.3). Right: Dust extinction map ($A_V$), derived
   comparing the measured H$\alpha$/H$\beta$ line ratio map with the
   calculated for case B recombination (left panel), and assuming the
   Cardelli, Clayton \& Mathis (1989) extinction law.}
 \label{fig:AV}       
 \end{figure*}

 Figure \ref{fig:Ha}, left panel, shows the derived H$\alpha$ flux map.  The
 well known structures of this region of the Orion nebula are easily
 identified: the strong H$\alpha$ emission area at the west of the Trapezium,
 the ``Dark Bay'', a strong decrease in the H$\alpha$ intensity, at the east
 of the Trapezium, classically explained by dust absorption, and the Orion
 ``Bar'', a filamentary area of strong H$\alpha$ emission at the south-east of
 the Trapezium. Other fainter structures, like the small arc of M42-HH2 at the
 north-west of the Trapezium can also be discerned. The different
 sub-structures seen in the reconstructed H$\alpha$ intensity map can be
 compared individually with those in the published HST and ground-based
 observations, which reinforce the reliability of our adopted procedure and
 results. In particular, this figure can be compared with the one derived from
 FP observations (Fig. 1e, \cite{pogg92}), since it does not include any
 contamination from [NII].  Despite the fact that there is an offset between
 the sampled areas of $\sim$1$\arcmin$ they agree perfectly well, indicating
 that our final resolution must be similar to the one obtained by them (i.e.,
 $\sim$2$\arcsec$).

 \begin{figure}
   \centering \centering \resizebox{\hsize}{!}
   {\includegraphics[width=7.8cm,angle=-90]{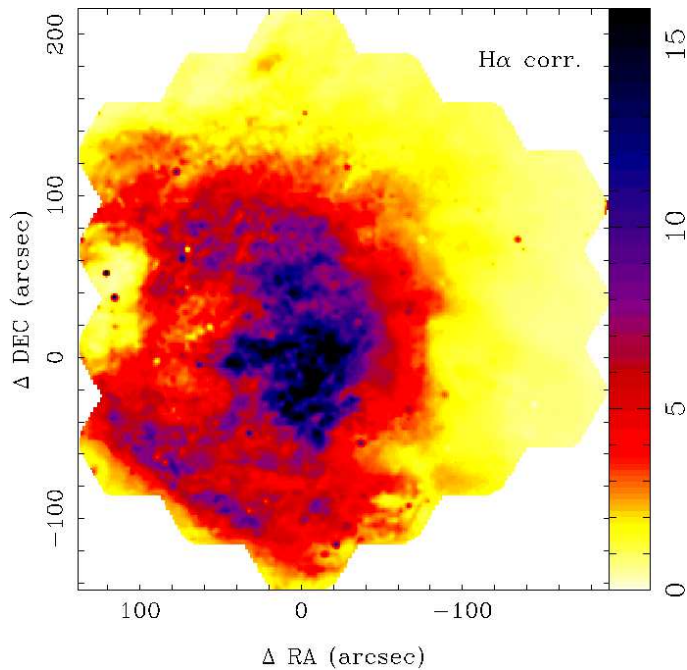}}
 \caption{H$\alpha$ intensity map corrected for dust extinction
in units of 10$^{-12}$ erg s$^{-1}$ cm$^{2}$ arcsec$^{-2}$ from the
uncorrected H$\alpha$ intensity map (Fig. \ref{fig:Ha}), and correcting by the
derived dust extinction map (Fig. \ref{fig:AV}).}
 \label{fig:Ha_corr}       
 \end{figure}

 Figure \ref{fig:Ha}, right panel, shows the observed H$\alpha$/H$\beta$ line
 ratio map. As stated before, this ratio can be used to trace the dust
 extinction when compared with the intrinsic one. The reddening of stars in
 the foreground of the Orion nebula is known to be low ($A_V\sim$0.2 mag;
 Breger, Gehrz \& Hackwell 1981), while the reddening towards the Trapezium is
 much higher ($A_V\sim 1.0-1.4$ mag; \cite{oster92}; Bohlin \& Savage 1981).
 Therefore most of the extinction is associated with the nebula. Pogge et al.
 (1992) proposed that it must arise in the neutral gas between us and the
 ionized zones of the nebula. Therefore, it is possible to assume a simple
 screen geometry for the dust extinction. This assumption implies that most of
 the deviations of the H$\alpha$/H$\beta$ ratio from the expected ones are due
 to dust. Pogge et al. (1992) estimated an average intrinsic ratio of
 H$\alpha$/H$\beta$=2.86 throughout the nebula, assuming case B recombination
 for a electronic temperature of T$=$9000 K and a density of $N_e\sim$10$^3$
 cm$^{-3}$, to derive their extinction map. Instead of using this
 simplification, we are able to determine both the electron density and
 temperature maps, and derive the expected H$\alpha$/H$\beta$ line ratio map
 for case B recombination.

 \begin{figure*}
   \centering \centering \resizebox{\hsize}{!}
   {\includegraphics[width=7.8cm,angle=-90]{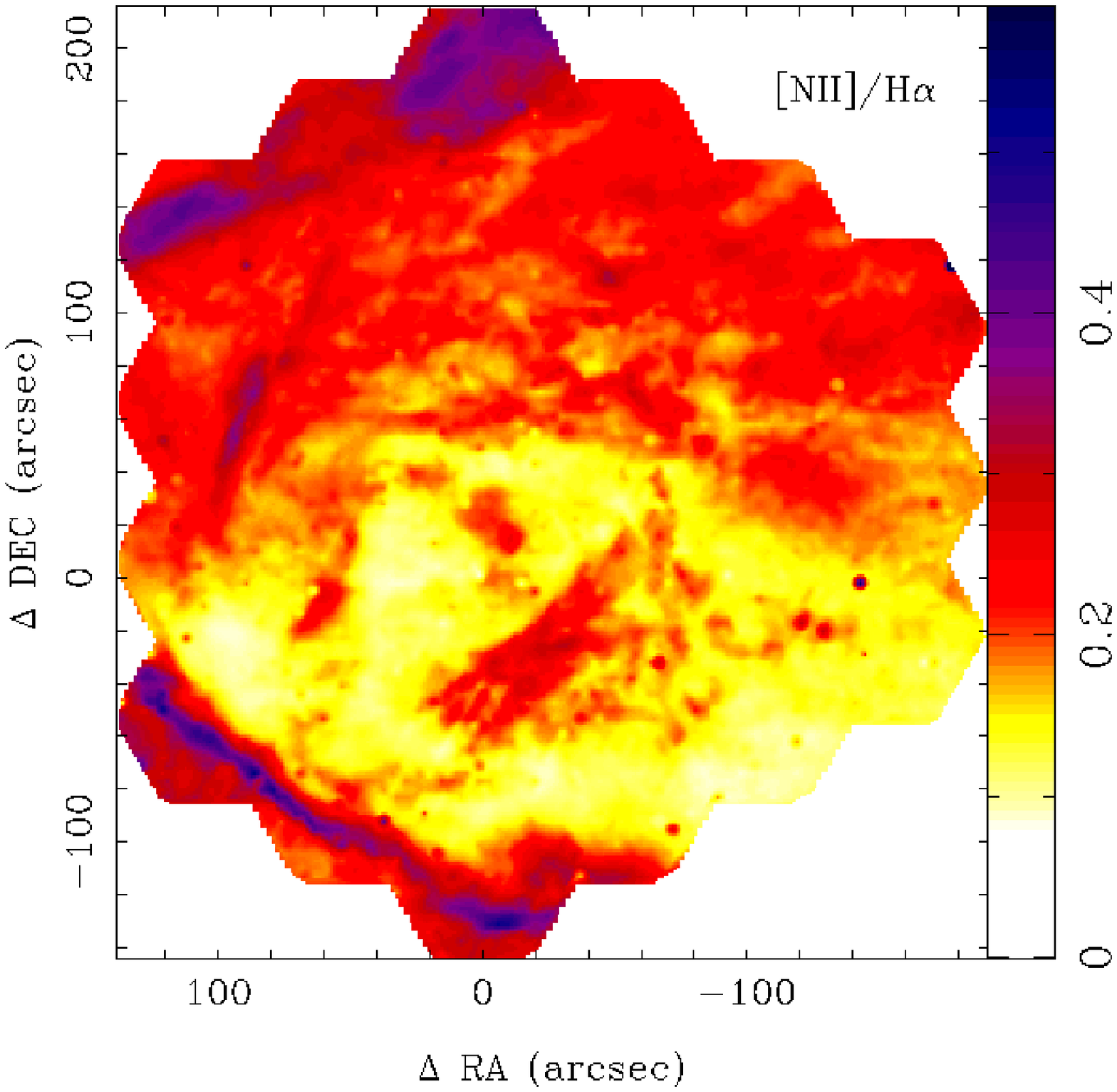}\includegraphics[width=7.8cm,angle=-90]{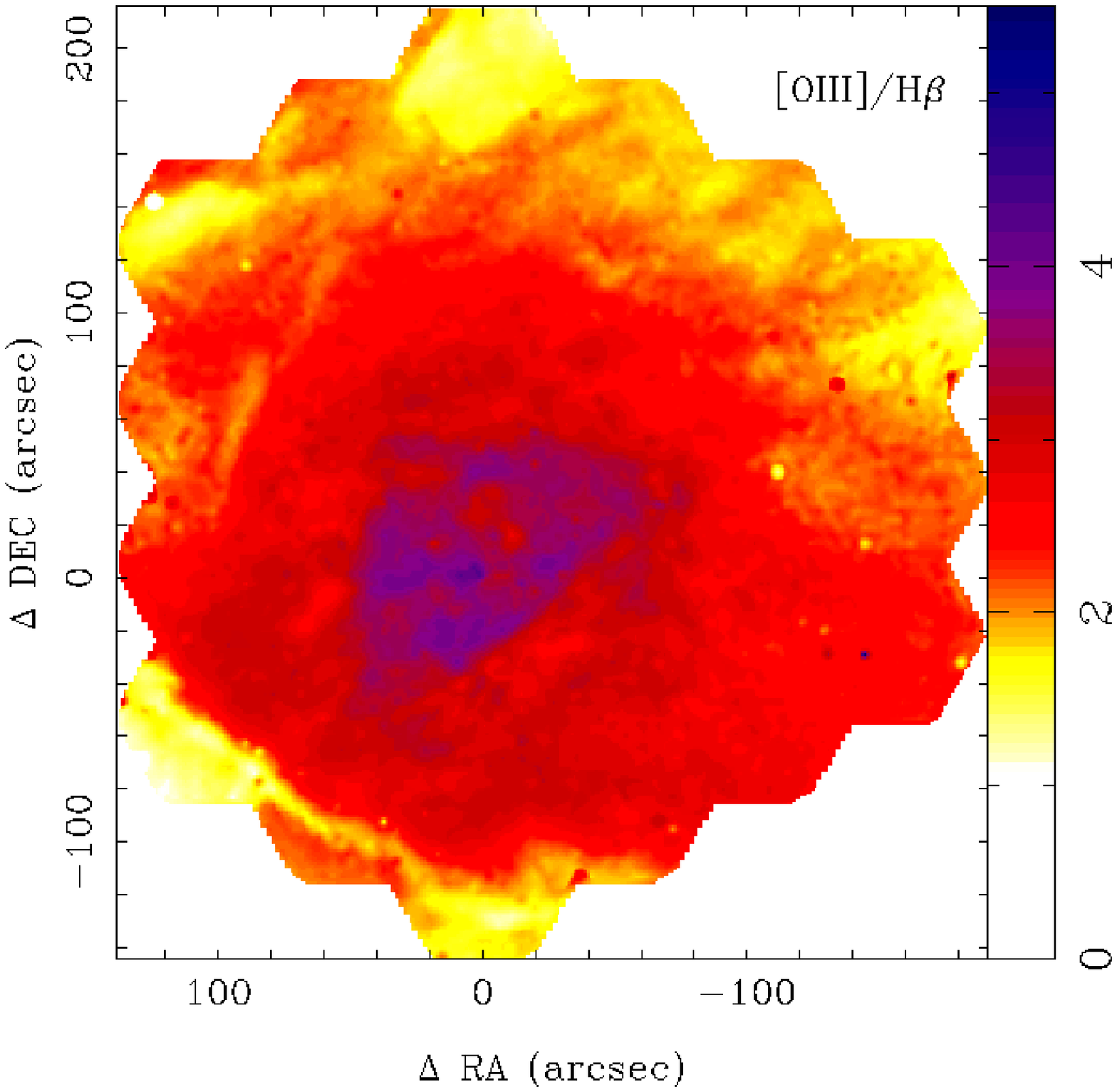}}
 \caption{Classical diagnostic line ratio maps. Left: [NII]$\lambda$6583/H$\alpha$ line ratio
   map. Right: [OIII]$\lambda$5007/H$\beta$ line ratio map.}
 \label{fig:line_ratio}       
 \end{figure*}

\subsubsection{Electron density and temperature}

As already indicated, it is possible to determine the electron density and
temperature using the [SII]$\lambda$6716,6731 doublet ratio and the
[NII]$\lambda$6548+83/[NII]$\lambda$5755 ratio. Figure \ref{fig:Ne_T} shows
the electron density map, $N_e$ (left panel) and the electron temperature map
(right panel) derived from the indicated line ratio maps, following the
formulae by Osterbrock (1989), as indicated in Section \ref{sec:inte}.  The
estimated density represents only a lower limit to the real one in the
hydrogen ionized zone, since the [SII] emission is also present in partially
ionized regions, especially near the ionization fronts, where the density can
change.

To our knowledge this is the first time that both the electron density and
temperature maps are estimated simultaneously. Pogge et al. (1992) derived an
electron density map using the [SII]$\lambda$6716,6731 doublet ratio but
assumed an average temperature of T=9000 K. This approximation is valid, if we
take into account the reduced effect of the temperature in the density
estimation and the narrow ranges of estimated temperatures throughout the
nebula (Fig.  \ref{fig:Ne_T}). This explain why, despite the differences in
the procedure to derive both maps, a considerable agreement in the shape of
the substructures and the covered range of electron densities is found between
both studies ($N_e\sim$100-15000 cm$^{-3}$). This agreement is expected since
the derivation of the electron density does not depend strongly on the
temperature. However, we find a richer level of detail in the substructures in
our map than that of Pogge et al. (1992). Since the final spatial resolutions
of both datasets are similar, we may assume that these substructures are real,
most probably induced by the mild effect of the temperature in the density
derivation.

The highest densities are found in two regions: (1) two clumps nearby and
towards the west of the Trapezium and (2) a set of clumps at $\sim$60$\arcsec$
south-southwest of the Trapezium, both regions with an electron density of the
order of $N_e$$\sim$3$\times$10$^4$ cm$^{-3}$. The first region was not
noticed by Pogge et al. (1992), since it was too near to their masked areas
due to the contamination of the bright stars of the Trapezium.  The second one
is coincident with the observed peak in the radio continuum emission
(Yusef-Zadeh 1990), and the [CII] $\lambda$ 158 $\mu$m IR structure detected
by Stacey et al. (1992), as already noticed by Pogge et al. (1992).  Our
density map is richer in substructures, and most of them are coincident with
the higher resolution structures in this region. It seems that this is a dense
region on the backside of the nebula (Pogge et al.  1992).

An enhancement of the density occurs in the Orion bar ionization front, from
$\sim$2000 cm$^{-3}$ to $\sim$4000 cm$^{-3}$, and then it drops again across
the front from north-west to south-east. The regions of higher density and the
detected substructures are also coincident with the areas of higher radio
continuum emission in the bar (Yusef-Zadeh 1990). This is consistent with the
classical picture of ionization fronts described by Hester (1991). In addition
to these large structures, there are a number of isolated substructures that
can be easily associated with known HH and HH-like objects in the nebula
(Pogge et al. 1992), although it is out of the scope of the current study to
describe or analyze them in detail.

As we indicated before, several previous spectroscopic studies have attempted
to determine the ionization conditions in the Orion nebula by obtaining slit
spectroscopy at different locations (e.g., \cite{bald91}; \cite{oster92};
\cite{este98}; \cite{este04}). Most of them have determined the electron
density and temperature at these locations using the same line ratios (usually
S$^{++}$ and N$^{++}$) and formulae used here. Baldwin et al.  (1991) obtained
low-resolution optical slit spectroscopy with the slit located at
$\sim$30$\arcsec$ west of $\Theta^1$ Ori C, in an east-west orientation,
reaching $\sim$5$\arcmin$ west.  In the range covered by our IFS data they
found a decay of the electron density from $\sim$6310 cm$^{-3}$ to $\sim$158
cm$^{-3}$ and of the electron temperature from $\sim$11000 K to $\sim$8000 K.
At the same location we found similar values, with the electron density
dropping from $\sim$6166 cm$^{-3}$ to $\sim$166 cm$^{-3}$, and the temperature
from $\sim$10000 K to $\sim$ 7000 K. Esteban et al. (1998) obtained high
resolution echelle spectroscopy at two locations of the nebulea, one at
45$\arcsec$ north and another at 20$\arcsec$ south and 10$\arcsec$ west of
$\Theta^1$ Ori C. New deeper data at a similar resolution were obtained by
Esteban et al. (2004) in the second location. They found an electron density
of 4020 cm$^{-3}$ and 6410 cm$^{-3}$, and a electron temperature of 9850 K and
10710 K at each of these locations. Using our IFS data we found an electron
density of $\sim$3716 cm$^{-3}$ and $\sim$7141 cm$^{-3}$ and a temperature of
$\sim$9743 K and $\sim$10217 K at the same locations. Osterbrock et al. (1992)
obtained optical and near-infrared low-resolution spectroscopy at
$\sim$45$\arcsec$ north of the Trapezium. They estimated the electron density
and temperature to be $\sim$4000 cm$^{-3}$ and $\sim$9000 K, similar to the
values derived using our IFS data at a similar location.

O'Dell et al. (2003) presented a high resolution map of the electron
temperature based on the [OIII] line ratio described above derived from
narrow-band imaging using the WFPC at the HST. Their narrow-band images have
the characteristic WFPC field-of-view of
$\sim$160$\arcsec$$\times$160$\arcsec$, located to the south-west of
$\Theta^1$ Ori C. Despite their very fine resolution, the presented image has
a not much better signal-to-noise than the one derived using our IFS data.
The reason for that comes from the steps required to derive the electron
density from narrow-band imaging, and how they affect the signal-to-noise of
the final derivation. They used a different ion to derive the electron
temperature (Fig. \ref{fig:Ne_T}). However we can compare them since there is
a simple relation between both temperatures (e.g. Pilyugin et al.  2006). The
values presented by O'Dell et al. (2003) in their Fig. 2 at different selected
locations and the average temperature throughout the field-of-view roughly
agree with those obtained using the presented IFS data at similar locations.
The temperature distribution is also coincident, however their lower
signal-to-noise ratio prevent us from performing a more detailed comparison.
The authors claim that although they found real variations of the temperature
across the nebula, based on a statistical analysis, there is no obvious
relation to the position in the nebula. Based on the temperature distribution
shown in Fig.  \ref{fig:Ne_T} we cannot agree with their interpretation of the
data, since we found an obvious rich structure in the temperature
distribution, which rises near the Trapezium and drops in the outer
regions of the field-of-view of our data.

Thus, the electron densities and temperatures derived using the presented IFS
data agree with those found using classical slit spectroscopy and narrow-band
imaging at different locations in the nebula.

\begin{figure*}
   \centering \centering \resizebox{\hsize}{!}
   {\includegraphics[width=7.8cm,angle=-90]{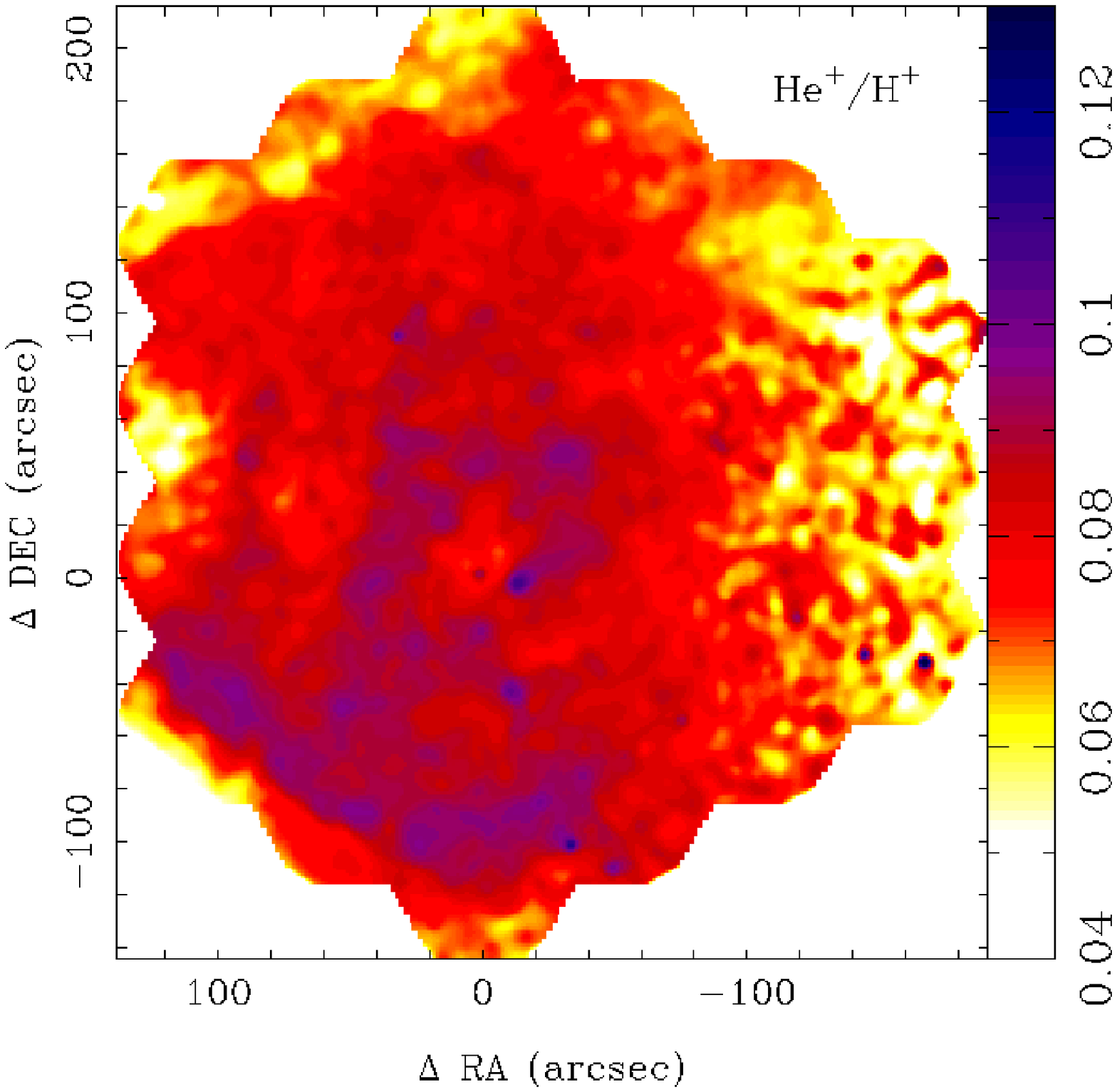}\includegraphics[width=7.8cm,angle=-90]{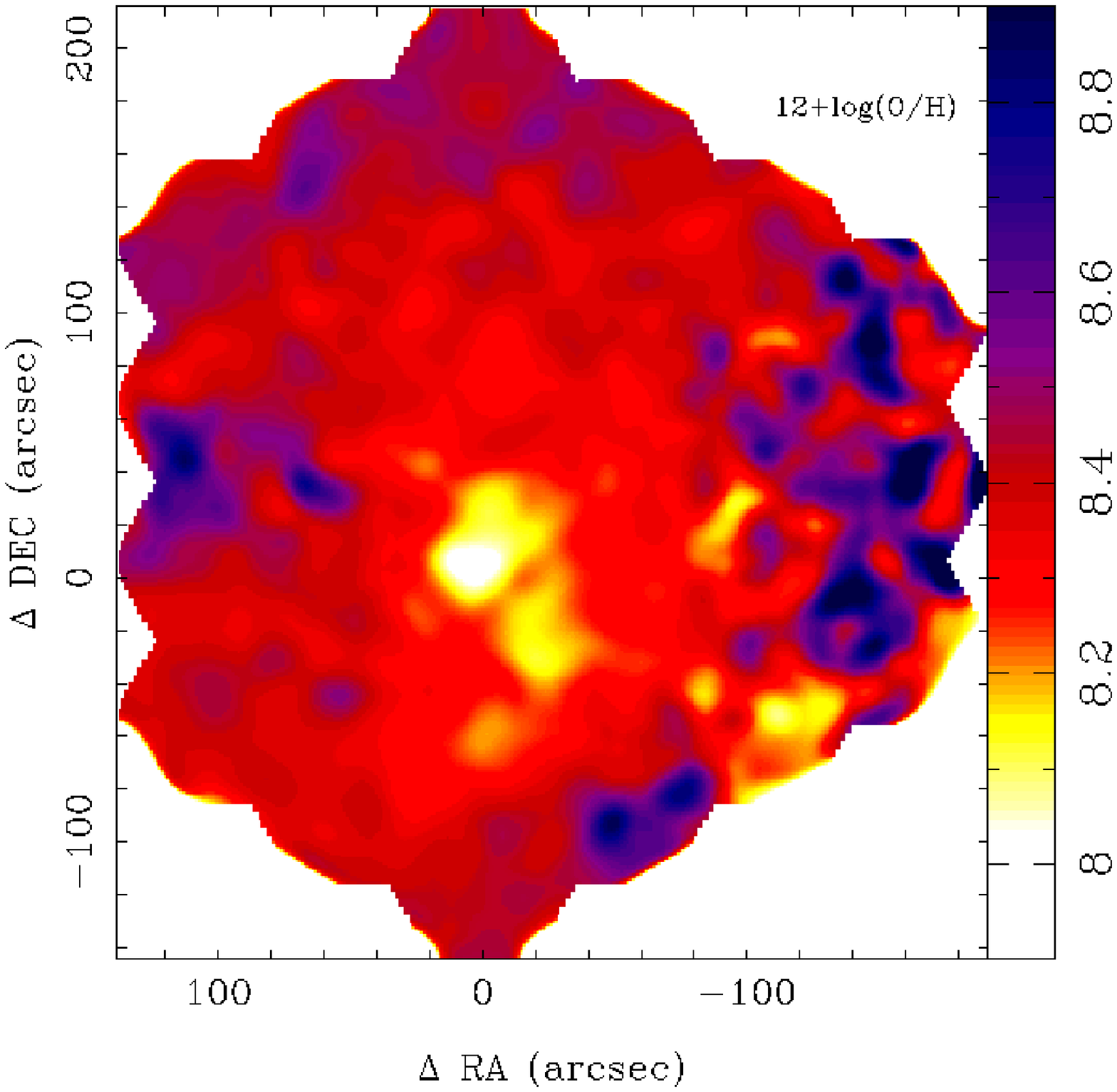}}
   \caption{Left: Map of the ionic abundance of He$^+$/H$^+$ derived from the
     H$\lambda$6678/H$\alpha$ line ratio map corrected for the effects of
     extinction and collisional excitation. Right: Oxygen abundance map
     derived directly using the measured temperature values and intensities of
     the [OII] and [OIII] lines.}
 \label{fig:He}       
 \end{figure*}

\subsubsection{Dust extinction}

Figure \ref{fig:AV}, left panel, shows the expected distribution of the
H$\alpha$/H$\beta$ line ratio map calculated for case B recombination using
the electron density, $N_e$, and temperature maps described above (Fig.3). We
have employed the line ratios listed in Osterbrock (1989) for a discrete set
of electron densities and temperatures in order to create a grid whose
interpolation allowed us to estimate the line ratio for any possible set of
these values. This estimation is just an approximation and a more accurate one
would require a complete modeling of the nebula. However, it is clearly a much
better approximation than any one used in previous studies, where a single
intrinsic H$\alpha$/H$\beta$ line ratio was estimated for the entire nebula
(e.g., \cite{oster92}; \cite{pogg92}). The estimated intrinsic line ratio
ranges between $\sim$2.8 and $\sim$3.3, showing a rich structure, with a clear
minimum nearby the Trapezium, towards the south-east, and a second one, less
prominent, at $\sim$40$\arcsec$ south-west. Most of the substructures are
clearly associated with fluctuations in the electron density and/or
temperature (see Fig. \ref{fig:Ne_T}).

Figure \ref{fig:AV}, right panel, shows the extinction map ($A_V$) derived
using the observed H$\alpha$/H$\beta$ line ratio map presented in Fig.
\ref{fig:Ha} and assuming the estimated intrinsic value of this line ratio for
each position described above (Fig.  \ref{fig:AV}, left panel). We used the
Orion extinction law of Cardelli, Clayton \& Mathis (1989), with a ratio of
total to selective extinction of $R_V=$5.5 (\cite{math81}), as in previous
sections.  In some locations the observed line ratio is lower than the
estimated one. As we indicated previously, case B recombination is the best
simple approximation to describe the physical conditions in this nebula, but
it does not completely describe the ionization conditions. Indeed, the
presence of dust mixed with the emitting gas decreases the real value of the
H$\alpha$/H$\beta$ ratio by a few percent. This deviation is most probably the
cause of the observed discrepancy between the estimated intrinsic line ratio
and the observed one in areas with low dust extinction.
 
Therefore the dust extinction map shown in Fig. \ref{fig:AV} has to be
considered as a lower limit to the real one, since the intrinsic line ratio
may be overestimated. The described effect is most probably not homogeneous,
which prevents us from applying a single correction to the full dataset. A
detailed modeling of the ionization conditions would be required to perform a
better estimation of the spatial distribution of the dust extinction. Such a
modeling is out of the scope of the current study.

As expected, the ``Dark-Bay'' structure is associated with an increase in the
foreground dust extinction, with two extensions towards the north and the
south of the Trapezium. Another peak of the extinction is located 2 arcmin
south-west of $\Theta^1$ Ori C. Similar dust extinction distributions were
reported by Pogge et al. (1992) and O'Dell et al. (2000). The Pogge et al.
(1992) extinction map was affected by ring-like reduction artifacts, typical
in Fabry-Perot data, absent in the currently presented dust extinction map.
O'Dell et al. (2000) derived the extinction coefficient ($C_{H\beta}$) using
two different methods: (i) by comparing high resolution radio continuum
emission maps obtained with the VLA and WFPC/HST H$\beta$ narrow-band images
and (ii) by deriving the H$\alpha$/H$\beta$ line ratio using WFPC/HST
narrow-band images. In both cases it was assumed that the intrinsic
H$\alpha$/H$\beta$ line ratio was constant across the nebula. Despite this
different approach their high resolution extinction maps are similar to the
one presented here, although the estimated values for the extinction seem to
be slightly lower for the IFS data.

 Figure \ref{fig:Ha_corr} shows the extinction-corrected H$\alpha$ flux map,
 after the dust extinction map shown in Fig. \ref{fig:AV} has been applied. As
 already noticed by Pogge et al.  (1992), the appearance of the nebula is not
 significantly modified by the structure in the foreground dust extinction.
 The most remarkable structure unaffected by the dust extinction correction is
 the Orion bar. The decrease of the H$\alpha$ emission south-east of this
 structure is not due to dust extinction, but rather a flattening of the
 surface of the nebula (\cite{wen95}). The ``Dark Bay'' is less
 prominent after correcting for dust extinction, but it is still present,
 indicating that it is not simply a region of high foreground extinction but a
 region with an absence of emission. The fact that this structure is
 also seen in radio continuum and CO maps (e.g., \cite{peim88}; \cite{werf89};
 \cite{thro86}) strongly supports this conclusion.

\subsubsection{Ionization structure}

The ionization structure of the nebula can be in principle investigated by
exploring the line ratio maps for the usual diagnostic lines (e.g.,
\cite{bald81}; Osterbrock 1989). Figure \ref{fig:line_ratio} shows the line
ratio maps for [NII]/H$\alpha$, left panel, and [OIII]/H$\beta$, right panel.
As already noticed by Pogge et al. (1992), the first one is the richest in
structure \footnote{The [OIII]/H$\beta$ map presented by Pogge et al. (1992)
  has an error of a factor 10 in the displayed scale, easy to cross-check
  when comparing with the values reported in their Table 2.}. The Orion bar is
the most prominent substructure, clearly seen as a diagonal region south-east
of the Trapezium. Another two structures are also seen as an enhance of the
[NII]/H$\alpha$ ratio, one a less prominent north-south bar to the east of the
Trapezium, at the edge of the ``Dark Bay'', and another one at the north-east
end of the field-of-view of our IFS data.  The [NII] emission is originated in
the single ionized regions between the ionized and partially ionized zones in
the nebula, tracing the rapid changes in local ionization, while [OIII] is
originated in the fully ionized zones (e.g., Osterbrock 1989), tracing the
level of ionization in these zones. The [NII]/H$\alpha$ as well as
[OII]/H$\beta$ increases in areas where the [OIII]/H$\beta$ drops, thus the
maps show the effective reciprocity between both ratios.

Pogge et al. (1992) noticed that it is not possible to use the
classical diagnostic diagrams (e.g., [OIII]/H$\beta$ vs. [NII]/H$\alpha$) to
distinguish between regions dominated by photoionization (HII-regions like)
and regions dominated by shocks. We have repeated the experiment using our own
data. They confirm that the line ratios at any position in the nebula are
located in the general locus of HII-regions. No major deviations
are found, even in the location of well known HH objects, like M42-HH2 and
M42-HH1, where shocks may play an important role. The reason is that these
regions are embedded in a high surface brightness photoionization nebula.  It
is known that many HH objects differ by $\sim$ 60 km s$^{-1}$ from the
systemic velocity of the nebula (e.g., \cite{cant80}; \cite{meab86}). Thus, it
is not possible to decouple them from the resolution of our dataset, requiring
high spatial and spectral resolution spectroscopy.


\subsubsection{He$^+$ Ionic abundance and metalicity}

The Orion nebula has been used as an anchor point for HII regions to determine
the primordial helium abundance ($Y_p$) and the slope of the relationship
between the helium and heavy-element abundance enrichment ($dY/dZ$), (e.g,
\cite{peim76}; \cite{bald91}). However little effort has been made to explore
possible internal effects on both quantities. Pogge et al. (1992) presented a
study of the systematic behavior of the helium emission with respect to the
hydrogen to understand the effects that influence the derived helium
abundance.  They presented the distribution of this abundance (He$^+$/H$^+$
map) in comparison to the ionization structure in the nebula.

Following Pogge et al. (1992) we determined the ionized helium abundance based
on the HeI$\lambda$6678/H$\alpha$ line ratio. In ionized nebulae it is
expected that the He$^+$ zone is spatially included in the H$^+$ one, and
therefore the relative strength of HeI to HI will provide a direct estimate
of the ionic abundance He$^+$/H$^+$ ($y^+$). To perform such a conversion we
used the effective recombination rates of Hummer \& Store (1987) for HI and of
Smits (1991) for HeI, recently updated by Benjamin et al. (1999,2002).
Figure~\ref{fig:He}, left panel, shows the derived He$^+$/H$^+$ ionic
abundance map, after applying a 4$\arcsec$$\times$4$\arcsec$ median filter to
increase the signal-to-noise ratio. As already noticed by Pogge et al. (1992),
there is a considerable degree of structure in the helium ionic abundance,
strongly correlated with the ionization structure in the nebula. In
particular, the partially ionized zones coincide with areas with lower
He$^+$/H$^+$ abundance. In these regions, the fraction of neutral helium can
be important, making it necessary to correct the derive helium abundance by
this neutral fraction using an ionization correction factor (ICF, e.g.,
\cite{vilc89}).

Baldwin et al. (1992) estimated the He$^+$/H$^+$ ionic abundance at
$\sim$30-90$\arcsec$ west from $\Theta^1$ Ori C at $\sim$0.0876$\pm$0.0047
cm$^{-3}$. At a similar location we estimate this value in $\sim$0.089 using
our IFS data. Esteban et al. (1998,2004) estimated the electron density at the
two locations described before as 0.0856$\pm$0.0069 cm$^{-3}$ and
0.0893$\pm$0.0092 cm$^{-3}$. At similar locations, we estimated the electron
density as $\sim$0.085 cm$^{-3}$ and $\sim$0.089 cm$^{-3}$. The agreement
between our results and previous studies at different locations reinforce our
approach and demonstrate the quality of the presented dataset.

Figure \ref{fig:He}, right panel, shows the oxygen abundance map using the
classical direct derivation using the temperature estimator (\cite{page79}).
The oxygen abundance was derived using the [OII]$\lambda$3727/H$\beta$ and
[OIII]$\lambda$5007/H$\beta$, together with the distribution of electron
densities and temperatures describe before, and the formulae presented by
Pilyugin \& Thuan (2005). The described electron temperature, T$_2$, derived
using the [NII] line ratio, is representative of single ionized ions (like
N$^+$ or O$^+$). This temperature is different to the temperature
representative of double ionized ions, T$_3$ (like O$^{++}$). To derive it,
one would have to use the [OIII] line ratio, as described above (Sec. 4.1).
However, the strong contamination by a Hg sky emission line at
$\lambda\sim$4358\AA\ close to the [OIII]$\lambda$4363 line prevents us from
using it, particularly in the regions of faint [OIII] emission. Instead we
have employed the model-independent relation between T$_2$ and T$_3$ presented
by Pilyugin et al.  (2006). For comparision we also estimated the oxygen
abundance ratio using the $O3N2$ indicator (\cite{pett04}), based on the
[OIII]/H$\beta$ and [NII]/H$\alpha$ line ratios maps (corrected for dust
extinction). This indicator was calibrated to determine the integrated
abundance in a nebula and it is not clear that it would be valid for
point-to-point abundance estimations within a nebula itselft.  It has the
advantage that it does not require the use of the poorly determined electron
temperature or density, but the use of a high signal-to-noise line ratios.
Although we found a one-to-one relation between both abundances, we also found
two worrisome effects: (1) there is a strong scatter in the relation, of the
order of $\sim$1 dex, and (2) the $O3N2$ relation induces a strong correlation
between the oxygen abundance and any of the indicators of the degree of
ionization, like the [OIII]/H$\beta$ and [NII]/H$\alpha$ line ratios and the
ionization parameter ($U$). These two effects prevent us from using this
latter indicator, although the one derived using the temperature measurement
has a much lower signal-to-noise ratio.
 
The derived distribution of oxygen abundace was smoothed using a
4$\arcsec$$\times$4$\arcsec$ median filter to increase the accuracy in the
areas of lower signal-to-noise ratio. Despite this smoothing the abundance
shows also a rich structure. However, in this case there is no clear relation
to the structure of the distribution of the ionization diagnostic lines, being
more related to the distributions of the electron densities and temperatures
shown in Fig.~\ref{fig:Ne_T}. The oxygen abundance has a minimum near the
Trapezium, of the order of 12+log($O/H$)$\sim$8.2, increasing towards the
outer parts, where it reaches 12+log($O/H$)$\sim$8.6--8.8.  Although in some
regions its structure seems to be correlated with that of the helium
abundance, contrary to what is expected, this is not always the case.  It is
out of the scope of this article to study in detail the reasons for such a
lack of correlation. However it is probably related to the presence of
non-fully ionized areas where the fraction of neutral helium is not
negligible, and/or a deplection of the oxygen from the helium ioniazed areas.

It is well known that there are two ionization zones in the Orion nebula
(\cite{odel01}). The zone closest to the ionization front, partially ionized,
where the fraction of neutral helium is not negligible, and the zone closest
to the ionizing star, fully ionized. The former is associated with an enhanced
[NII]/H$\alpha$ ratio, while the latter is associated with an increase of the
[OIII]/H$\beta$ ratio. Thus, in the partially ionized zone it would be
necessary to recompute the fraction of ionized helium by applying an ICF.

\section{Summary and conclusions}

We have performed a low-resolution IFS survey covering the optical
wavelength range ($\sim$3700-7100\AA) within a field-of-view of
$\sim$5$\arcmin\times$6$\arcmin$ centered approximately in the Trapezium area
of the Orion nebula. The data, comprising 8121 individual spectra sampling
circular areas of 2.7$\arcsec$ each are distributed public as a PPAK
legacy project. To demonstrate the use of these data we have analysed
the integrated spectrum, detecting 82 individual emission lines, whose fluxes
were measured.  Based on these line intensities we derived the main
integrated spectroscopic properties of the ionized gas, including the average
dust extinction, electron density and temperature and the values of the
classical diagnostic line ratios. All of them agree with previously published
results based on long-slit and Fabry-Perot observations.

Line intensity maps for different emission lines were obtained to study the
complex structure of the ionized gas in the nebula. The derived H$\alpha$
intensity map, a tracer of the ionized hydrogen, was found to be similar to
that found using FP and narrow-band imaging, which supports the overall
observational and reduction procedure. The electron density and temperature
maps were derived using the [SII] line ratios and [OIII] and [NII] line ratios
respectively. A considerable amount of structure was found in both
distributions. We used this stricture to create a map of the expected
H$\alpha$/H$\beta$ line ratio, based in case B recombination. A comparison
between this expected line ratio and the observed one was used to obtain a
dust extinction map that was used to correct for reddening all in the derived
emission line maps.

The ionized structure was studied based on the classical diagnostic line
ratios [NII]/H$\alpha$ and [OIII]/H$\beta$. The distribution of helium and
oxygen abundances were derived based on the HeI$\lambda$6678/H$\alpha$ line
ratio and the temperature metallicity indicator (based on the [OII]/H$\beta$
and [OIII]/H$\beta$ line ratios). No clear relation was found between the
abundance distributions, contrary to what was expected from the study of the
integrated properties of known HII regions. There is a relation between the
helium abundance and the ionization structure, in the sense that partially
ionized areas are less abundant in helium and more abundant in heavy elements.
Two possible deviations from case B recombination theory in the Orion nebula
may explain this trend and the lack of a relation between both abundances: (i)
the fact that the nebula is not fully ionized, which leads to an
underestimation of the helium mass fraction due to the non neglectible
contribution of neutral helium, and (ii) the suspiction that there is dust
mixed with the emitting gas which absorbs the Lyman-line ionizing photons.

A more detailed modeling is required to correctly understand the ionizing
conditions in the nebula and their relations with the helium and heavy metals
abundances. We consider that the released dataset will be of a fundamental
importance to constrain such models.

\begin{acknowledgements}
 
  
  We thank the Spanish Plan Nacional de Astronom\'\i a program
  AYA2005-09413-C02-02, of the Spanish Ministery of Education and Science and
  the Plan Andaluz de Investigaci\'on of Junta de Andaluc\'{\i}a as research
  group FQM322.
  
  MV thanks Matthias Steinmetz and the instrumentation
  group at the AIP for facilitating the construction of the PPak module
  and its integration into PMAS, the spectrograph used for this survey. 

  We thank the referee, Prof. Dr.  O'Dell, who helped us to improve the
  quality of this article with his valuable comments and suggestions, and his
  detailed check of the flux calibration.

\end{acknowledgements}

\end{document}